\documentclass[twocolumn,aps,showpacs,amsmath,amssymb,floatfix,superscriptaddress]{revtex4}
\usepackage{graphicx}
\usepackage{dcolumn}
\usepackage{bm}
\usepackage{hyperref} 
\usepackage{latexsym}
\usepackage{float}

\begin{document}
\title{Majority Rule Dynamics in Finite Dimensions}
\author{P.~Chen}
\email{patrick@bu.edu}
\author{S.~Redner}
\email{redner@bu.edu}
\affiliation{Center for BioDynamics, Center for Polymer Studies, 
and Department of Physics, Boston University, Boston, MA, 02215}

\begin{abstract}
  We investigate the long-time behavior of a majority rule opinion dynamics
  model in finite spatial dimensions.  Each site of the system is endowed
  with a two-state spin variable that evolves by majority rule.  In a single
  update event, a group of spins with a fixed (odd) size is specified and all
  members of the group adopt the local majority state.  Repeated application
  of this update step leads to a coarsening mosaic of spin domains and
  ultimate consensus in a finite system.  The approach to consensus is
  governed by two disparate time scales, with the longer time scale arising
  from realizations in which spins organize into coherent single-opinion
  bands.  The consequences of this geometrical organization on the long-time
  kinetics are explored.
 
\end{abstract}


\maketitle

\section{Introduction}

\begin{figure}[!ht] 
 \vspace*{0.cm}
 \includegraphics*[width=0.375\textwidth]{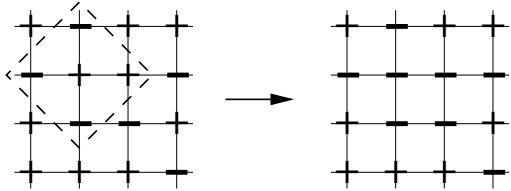}
\caption{Illustration of a single majority rule update step for 
  the 5-site von Neumann neighborhood on the square lattice.}
\label{process}
\end{figure}

The majority rule model (MR) is a simple description for consensus formation
in an interacting population.  The model consists of $N$ spins (opinions)
that are fixed on lattice sites, and each spin can assume the states $+1$ or
$-1$, corresponding to two opposite opinions.  Spins evolve by the following
two steps: first, pick a group of spins of fixed odd size $G$; second, all
the spins in this group adopt the state of the local group majority
(Fig.~\ref{process}).  These two steps are repeated until a final consensus
is necessarily reached.  Our goal is to understand basic properties of this
approach to consensus in finite spatial dimensions.

A general form of this majority rule dynamics was introduced by Galam
\cite{galam} in which a variable number of groups of arbitrary size are
formed simultaneously and then majority rule is simultaneously applied to
each group.  Our implementation of majority rule, in which only a single
small group is updated at each time step, allows for considerable analytical
progress in the mean-field limit \cite{MM,SL} and also makes it convenient to
simulate the model, especially in high dimensions.

In a previous study of the MR model \cite{MM}, it was shown that the average
time until consensus is reached is proportional to the logarithm of the
number of spins $N$ in the system in the mean-field limit.  On the other
hand, for finite dimensions, numerical simulations suggested that the most
probable consensus time grows as a power law in $N$, with an exponent that
decreases as the spatial dimension increases.  Mean-field behavior was not
reproduced even in four dimensions, indicating a still larger value for the
upper critical dimension of the MR model.

In this article, we focus on the MR model in finite spatial dimensions.  The
questions that we will investigate are: What is the geometry of
single-opinion domains?  How long does it take to reach consensus?  How do
basic system parameters affect the consensus time?  We find that the
probability distribution for the consensus time involves two very different
time scales when the spatial dimension is greater than one.  The longer time
scale arises from configurations in which opposite-opinion domains organize
into coherent geometries -- stripes in two dimensions, slabs in three
dimension, {\it etc}.  While the probability for the system to reach such a
coherent state decreases as the spatial dimension is increased --
approximately $33\%$ in two dimensions and $8\%$ in three dimensions -- we
believe that this probability remains non-zero in all finite spatial
dimension.  More importantly, the time needed to reach final consensus from
these coherent states is extremely long.  These configurations therefore give
the dominant contribution to the mean consensus time.

To put our results in context, it is instructive to compare the MR model with
two fundamental kinetic spin models, namely, the voter model (VM)
\cite{voter}, and the kinetic Ising model with zero-temperature Glauber
kinetics (IG) \cite{glauber}.  The VM describes consensus formation in a
population of individuals with zero self confidence.  In an update step of
the VM, a spin is selected at random and it blindly adopts the state of a
randomly-selected neighbor.  This step is repeated until consensus is
necessarily reached.  Because of the underlying linearity of the VM spin-flip
rate on the number of anti-aligned nearest neighbors, the VM is exactly
soluble in all spatial dimensions \cite{voter,pk,fpp}.  In particular, for an
$N$-spin system in $d$ dimensions with zero initial magnetization, the
consensus time scales as $N$ for $d>2$, as $N\ln N$ in $d=2$ (the critical
dimension of the VM), and as $N^2$ in $d=1$.  Because the average
magnetization is conserved, the probability that the system eventually ends
with all $+$ spins equals the initial density of $+$ spins in all spatial
dimensions.

In contrast, the zero-temperature kinetics of the IG model obeys a form of
majority rule.  In the update step, a flippable spin (those with zero or
positive energy) is picked at random and it adopts the state of the majority
in its interaction neighborhood.  In the case of a tie in the neighborhood
state (which can happen on bipartite lattices), the selected spin flips with
probability 1/2.  This elemental update step is repeated until no flippable
spins are left.  At early times, coarsening domains form whose typical length
scale grows as $t^{1/2}$ due to an underlying diffusive dynamics \cite{rev}.

The primary operational difference between the IG and MR models is that in
the latter {\em all\/} the spins within the neighborhood flip, a feature that
also occurs in Galam's model \cite{galam} and also in the Sznajd model
\cite{szn} of social influence, where a small group that is in consensus can
influence other spins at the periphery of the group.  This distinction in the
update rule has fundamental consequences.  In the IG model, infinitely
long-lived metastable states can occur that consist of perfectly flat
interfaces in $d=2$, or states where all interfaces have zero net curvature
for $d\geq 3$ Ref.~\cite{SKR}.  In contrast, consensus is the only possible
final state in the MR model.  Nevertheless, both the zero-temperature IG
model and the MR model have anomalous kinetics because of the existence of
very long-lived transient states.

In Sec.~II, we present simulation results for the anomalous behavior of the
consensus time distribution and the two basic controlling time scales.  Then
in Sec.~III, we discuss the role of the long-lived coherent states that
dominate the asymptotic tail of the consensus time distribution.  A
qualitative argument for the lifetime of these states is given in Sec.~IV.
We conclude in Sec.~V.

\section{Consensus Time Distribution}

We first simulate the distribution of times until consensus is reached on
finite-dimensional hypercubic lattices with periodic boundary conditions.
Typically, we initialize each realization of the system to contain equal
numbers of $+$ and $-$ spins.  We choose the group size to be $G=3$ and
construct the group by selecting a spin at random and then randomly picking
two out of its $2d$ nearest neighbors.  This definition for a group has the
advantages of computational simplicity and a dimension-independent group
size.  Other definitions for a group, such as the von Neumann neighborhood of
Fig.~\ref{process} (the initial site plus its $2d$ nearest neighbors; group
size $G=2d+1$), lead to qualitatively similar results.

We then evolve each realization according to MR kinetics until consensus is
reached.  The quantities that we focus on are: (i) the distribution of
consensus times, $P_N(t)$, in an $N$-spin system with zero initial
magnetization, and (ii) the probability for a realization to reach a stripe
or a slab state, $S_N(m)$ (to be defined below), as a function of $N$ and the
initial magnetization $m$.

\begin{figure}[!ht] 
 \vspace*{0.cm}
 \includegraphics*[width=0.44\textwidth]{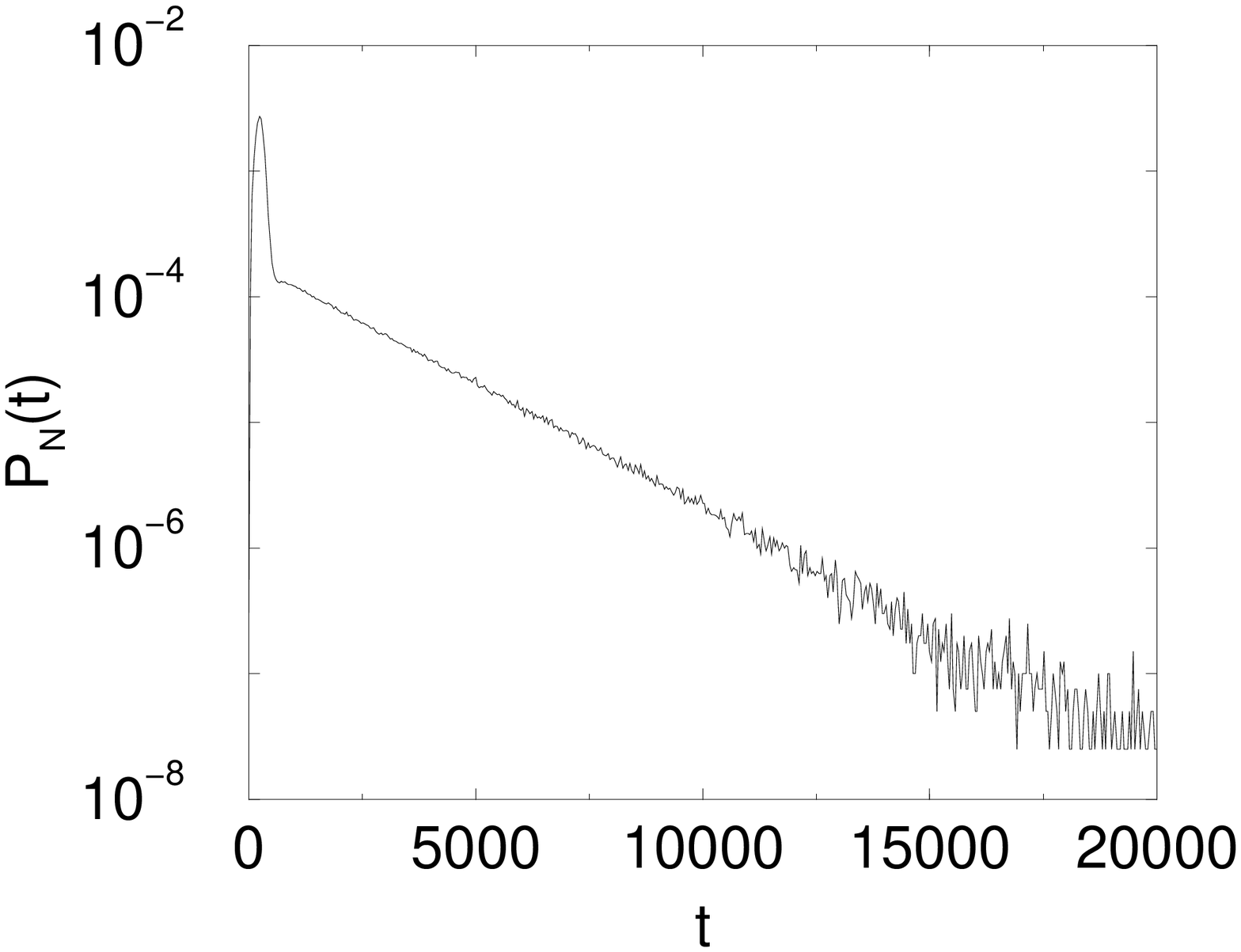}
 \includegraphics*[width=0.44\textwidth]{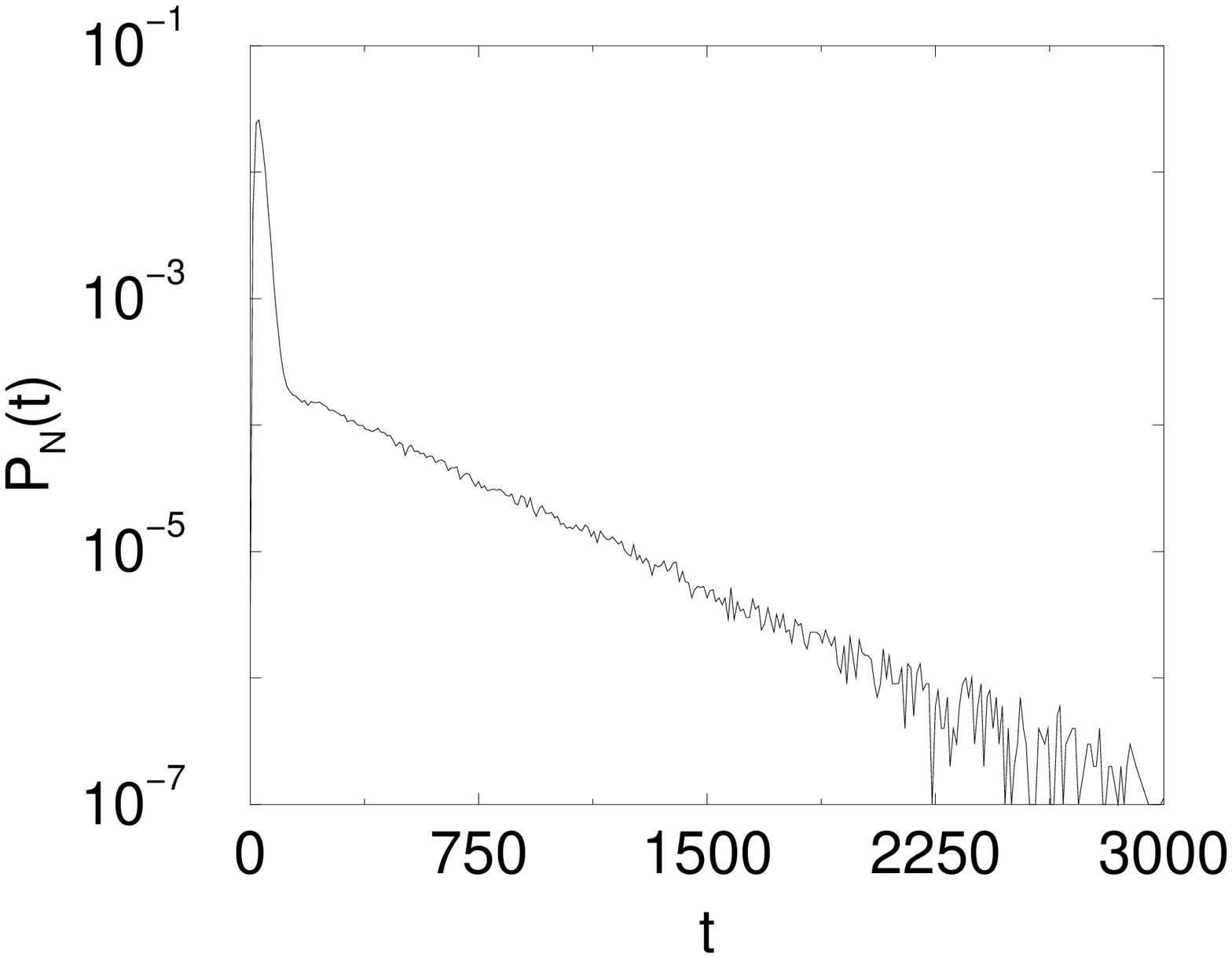}
 \includegraphics*[width=0.44\textwidth]{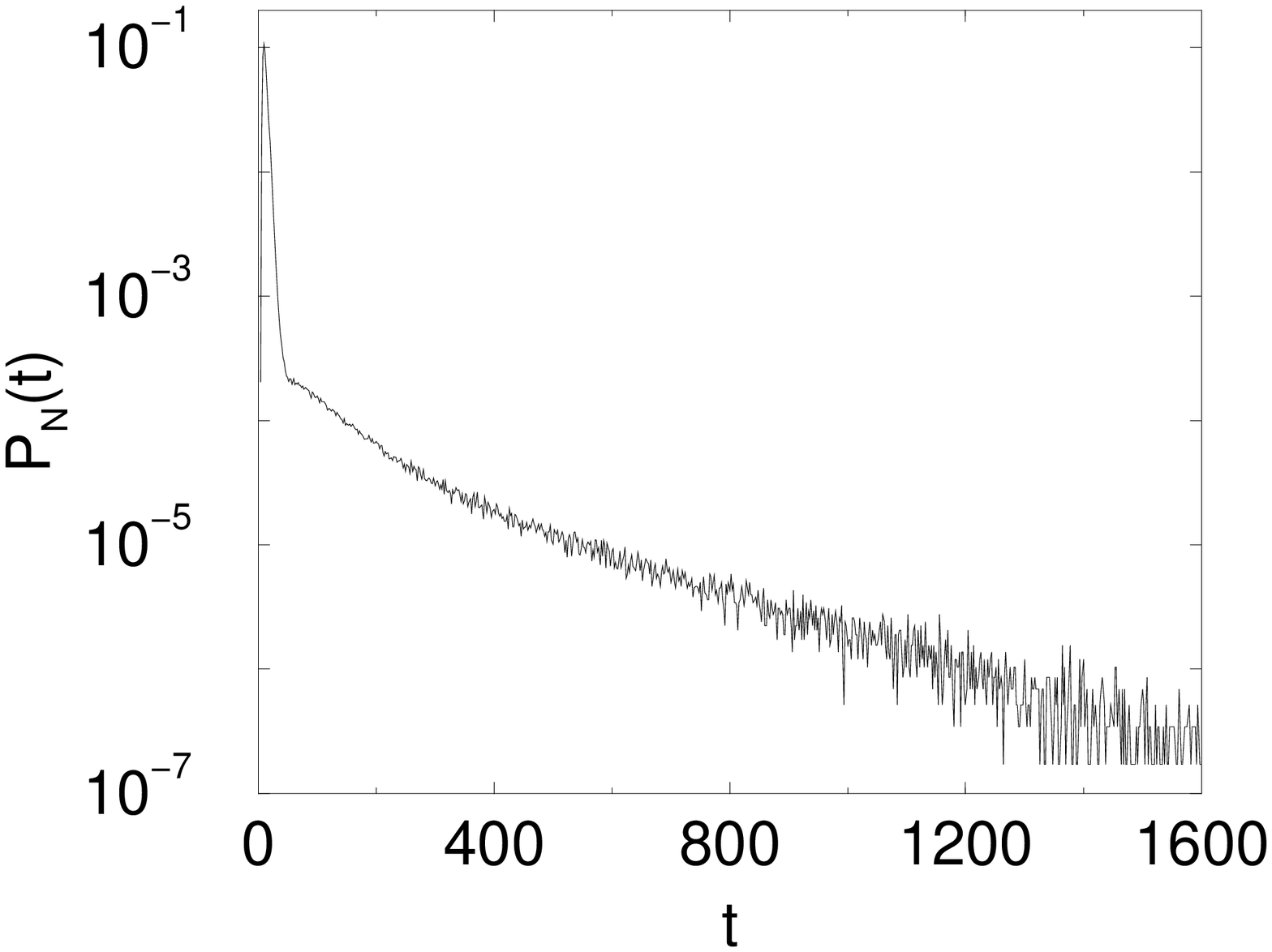}
 \caption{Distribution of consensus times, $P_N(t)$, versus $t$ (in units of
   Monte Carlo steps per spin) for an $N$-spin system starting from a random
   zero-magnetization state.  Shown are results for (top to bottom): a $50^2$
   square lattice ($N=2500$), a $14^3$ cubic lattice ($N=2744$), and a $7^4$
   four-dimensional hypercubic lattice ($N=2401$).  Data are all based on
   $10^6$ realizations.}
\label{PDF}
\end{figure}

The consensus time distributions $P_N(t)$ for spatial dimensions 2, 3, and 4
are shown in Fig.~\ref{PDF}.  It is evident that in two and three dimensions,
$P_N(t)$ is characterized by two time scales -- the most probable consensus
time, corresponding to the peak of the distribution, and a much longer time
scale associated with the asymptotic exponential decay.  In four dimensions,
there is a change in the slope of the asymptotic tail of $P_N(t)$ for $t\agt
400$, suggesting the possibility that the asymptotic kinetics involves yet a
third time scale.

\begin{figure}[ht] 
 \vspace*{0.cm}
 \includegraphics*[width=0.4\textwidth]{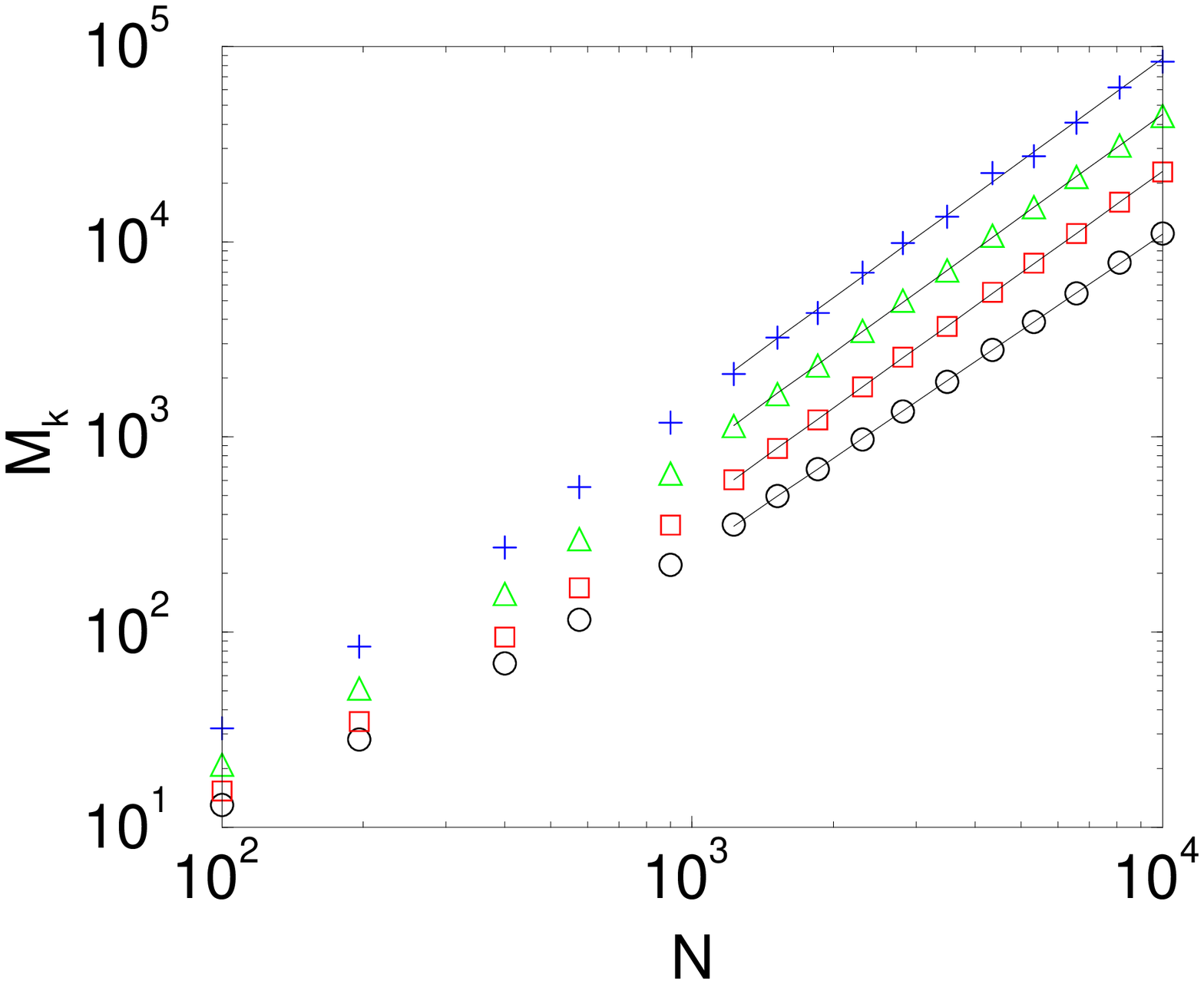}
 \includegraphics*[width=0.4\textwidth]{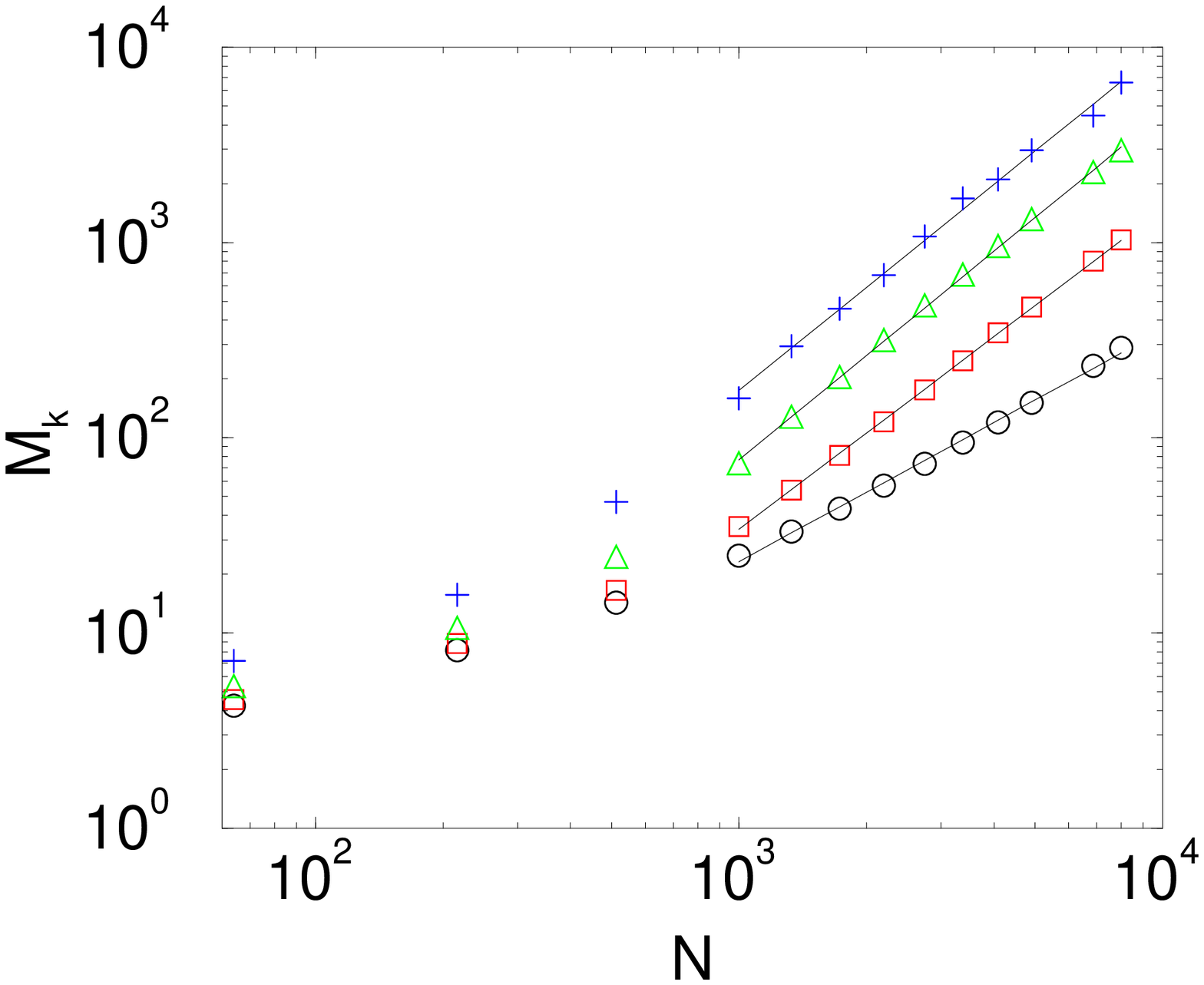}
 \caption{Double logarithmic plot of $M_k(N)$, the $k^{\rm th}$ reduced
   moments of the consensus time distribution, versus the total number of
   spins $N$ for $k=1$ ($\circ$), $k=2$ ($\Box$), $k=4$ ($\triangle$), $k=8$
   ($+$), in $d=2$ (top) and $d=3$ (bottom).  Data are based on $10^5$
   realizations in $d=2$ and $4 \times 10^5$ realizations in $d=3$. The lines
   are least-squares linear fits to the large-$N$ data.}
\label{Mk}
\end{figure}

From these data, we find that the most probable consensus time, $t_{\rm mp}$,
scales with $N$ as $t_{\rm mp}\sim N^\alpha$, with $\alpha\approx 1.24$,
$0.72$, and 0.56 for spatial dimensions 2, 3, and 4, respectively.  These
values are identical to those obtained previously in Ref.~\cite{MM}; these
were based on smaller-scale simulations in which realizations where the
consensus time exceeded a (large) preset limit were terminated.  

On the other hand, the asymptotic decay of $P_N(t)$ is clearly governed by a
much longer characteristic time and we now apply two methods to estimate this
longer time scale.  First, we consider the reduced moments of the consensus
time distribution
\begin{equation}
M_k(N) \equiv \langle (t(N))^k\rangle^{1/k} = 
\left[\int_{0}^{\infty} t^k\, P_N(t)\,dt\right]^{1/k}.
\end{equation}
As suggested by the data in Fig.~\ref{PDF}, if the long-time tail of
consensus time distribution has a simple exponential decay of the form
$e^{-t/\tau(N)}$ at long times, then all the reduced moments would
asymptotically scale as $\tau(N)$, with subdominant corrections that become
smaller as $k$ increases.  This trend is illustrated in Fig.~\ref{Mk} where
$M_k$ is plotted as a function of $N$ for various values of $k$.  In $d=2$,
each $M_k$ grows as a power law in $N$ for large $k$, but with a slightly
different apparent exponent.  Least-squares fits to the data give the
following exponents in $d=2$: 1.64 for $k=1$ ($\circ$), 1.73 for $k=2$
($\Box$), 1.75 for $k=4$ ($\triangle$), and 1.75 for $k=8$ ($+$).  From this
limiting large-$k$ value of this exponent we can then infer the
$N$ dependence of $\tau$.

For $d=3$, the behavior is qualitatively similar, except that there is a
large disparity in the exponents for $M_k$ for $k=1$ and for $k>1$.  Linear
fits to the data now give the exponent values 1.18 for $k=1$ ($\circ$), 1.64
for $k=2$ ($\Box$), 1.77 for $k=4$ ($\triangle$), and 1.78 for $k=8$ ($+$).
However, there is a perceptible downward curvature in the dependence of $M_k$
on $N$ for large $k$, so that linear fits are inadequate to determine the
$N$-dependence of $\tau$ accurately.

\begin{figure}[ht] 
 \vspace*{0.cm}
 \includegraphics*[width=0.415\textwidth]{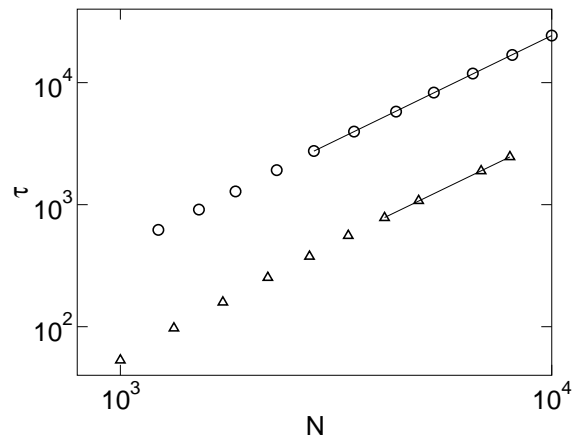}
 \caption{Double logarithmic plot of $\tau(N)$ (in units of Monte Carlo steps
   per spin) versus $N$ for $d=2$ ($\circ$) and $d=3$ ($\triangle$).  The
   lines are the best fits to the last few data points.}
\label{tau}
\end{figure}

Our second analysis method is simply to measure the slope of the exponential
tail of $P_N(t)$ directly for different values of $N$ and thereby determine
$\tau(N)$.  To do this, we first make a first estimate for $\tau$ by finding
the slope in the region that is visually most linear.  Then we refine this
estimate by computing the slopes in the systematic ranges $(\tau/2,2\tau)$,
$(\tau/2,3\tau)$, $(\tau/2,4\tau)$, {\it etc.}, and using the range where a
linear fit has the highest correlation coefficient.  In the resulting data
for $\tau$ versus $N$, there is now small and systematic downward curvature
(Fig.~\ref{tau}).  By dropping the first four data points one-by-one and then
performing linear fits to the remaining data, the local slope decreases from
1.746 to 1.719 in two dimensions.  In three dimensions, there is a larger
decrease in the local slope from 1.832 to 1.709 as the first six points are
deleted.  Extrapolating this local slope to $N\to\infty$, we obtain the
estimates $\nu= 1.7\pm 0.04$ in $d=2$ and $\nu=1.5\pm 0.1$ in $d=3$ in the
relation $\tau\sim N^\nu$.  The error bars are a subjective guess of the
uncertainty in the extrapolation.

\section{Anomalous Coarsening and Long-Lived Coherent States}

\begin{figure}[H] 
 \vspace*{0.cm}
  \includegraphics*[width=0.22\textwidth]{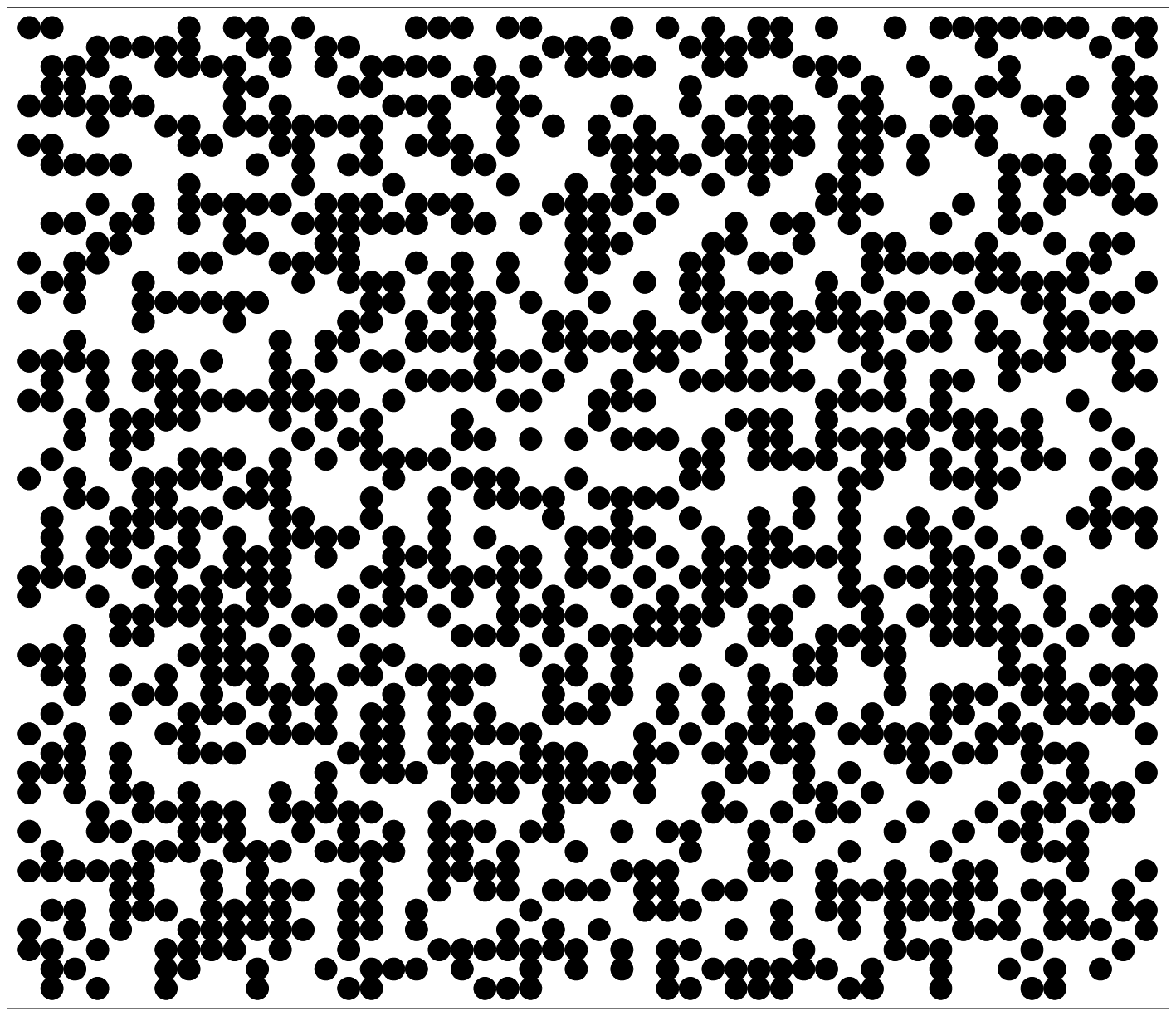}\hfill
  \includegraphics*[width=0.22\textwidth]{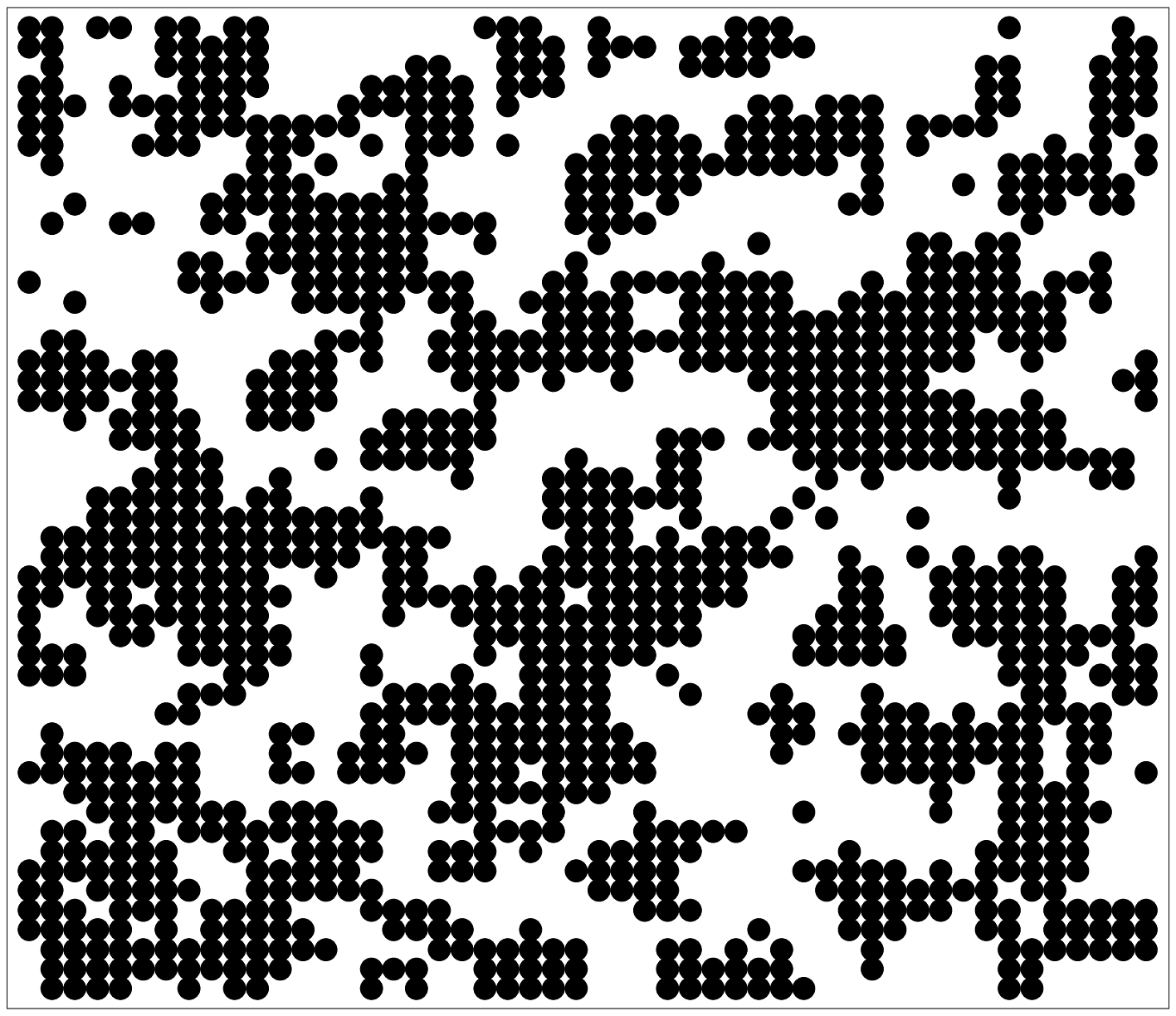}\\ \vskip -0.5ex
  \includegraphics*[width=0.22\textwidth]{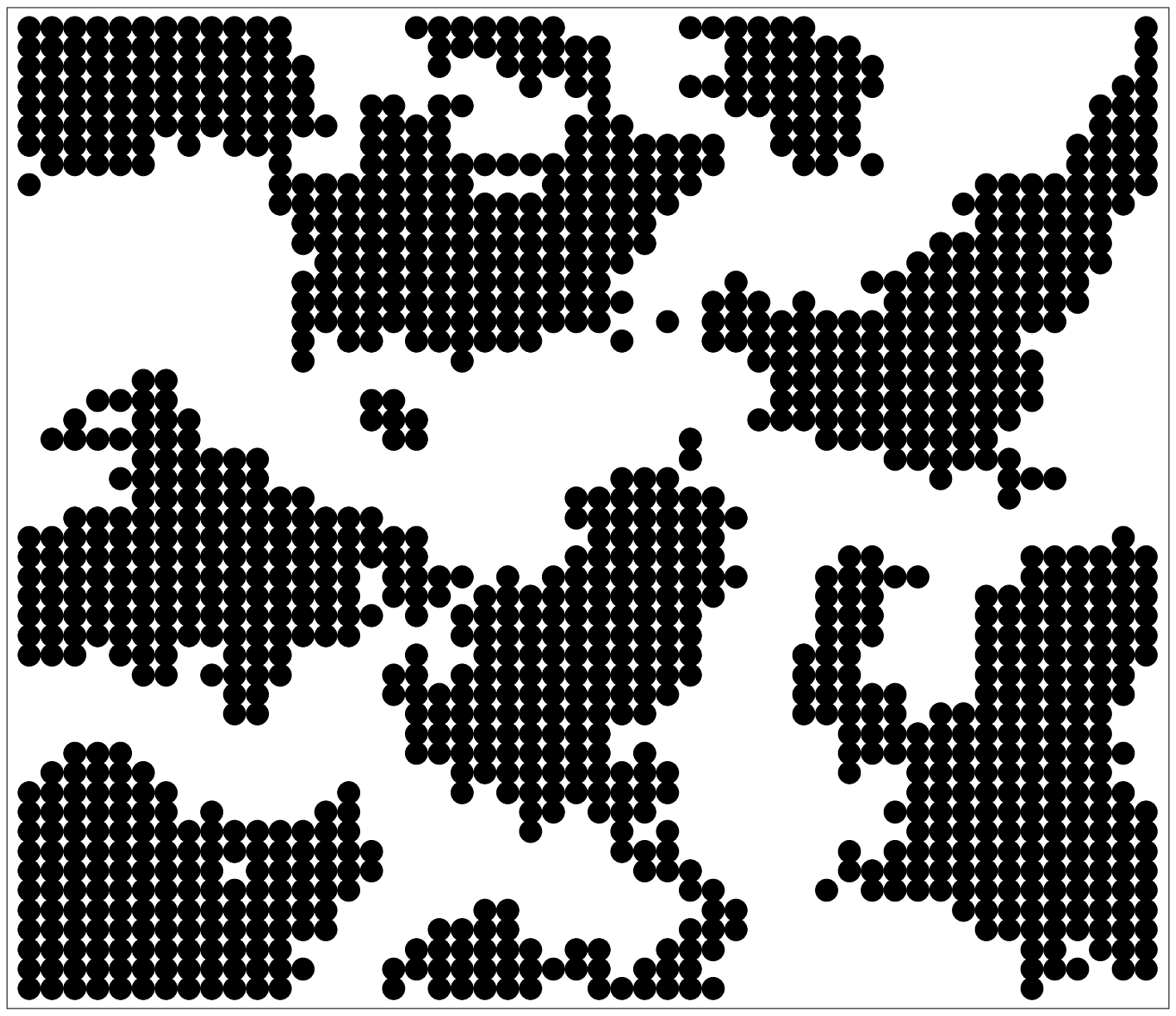}\hfill
  \includegraphics*[width=0.22\textwidth]{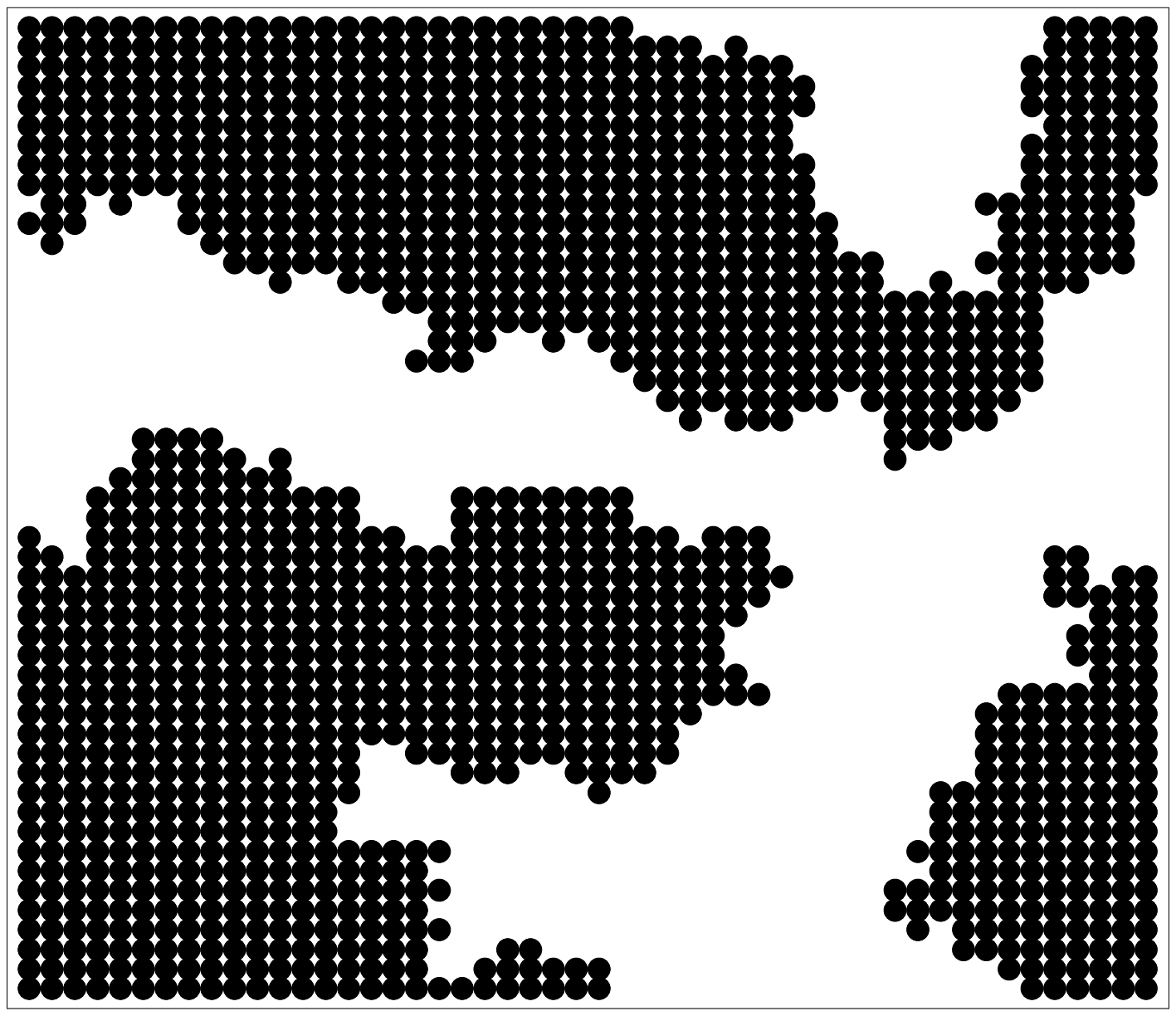}\\ \vskip -0.5ex
  \includegraphics*[width=0.22\textwidth]{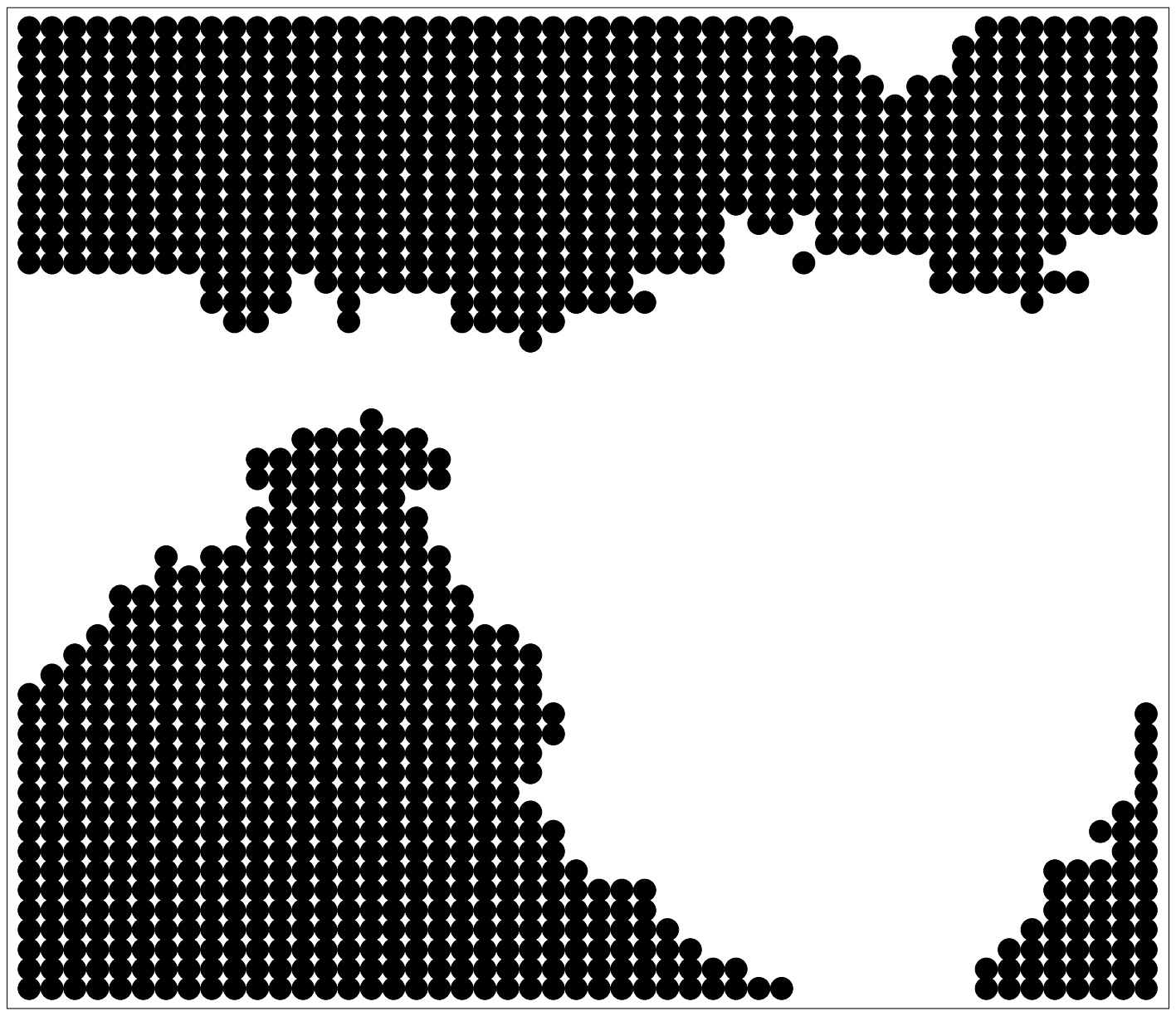}\hfill
  \includegraphics*[width=0.22\textwidth]{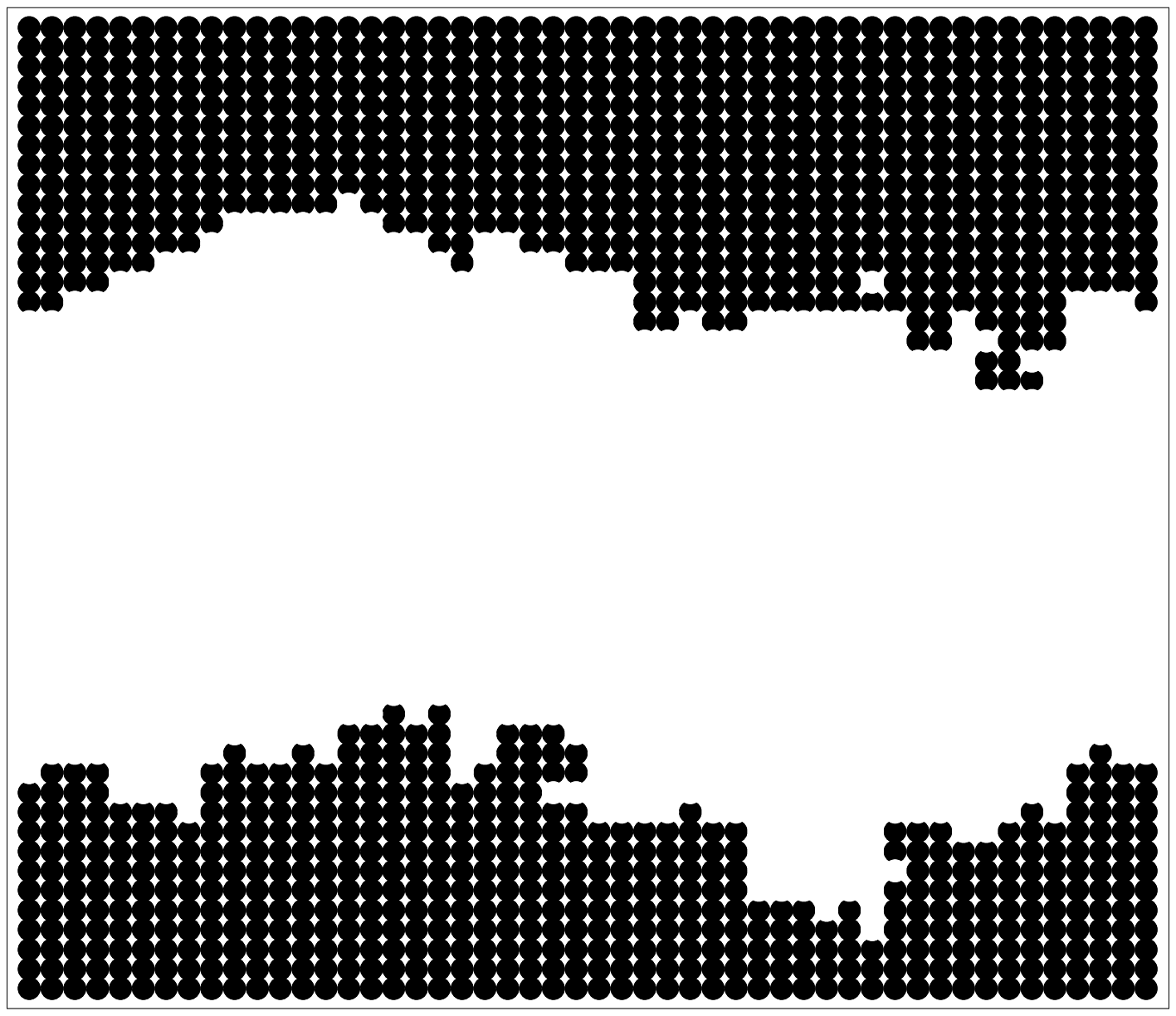} \\
\vspace*{0.2cm}
\caption{Snapshots of a $50^2$ system at $t=0,1,5,20,80$, \& 200.}
\label{pics}
\end{figure}

\begin{figure}[ht] 
 \vspace*{0.cm}
  \includegraphics*[width=0.5\textwidth]{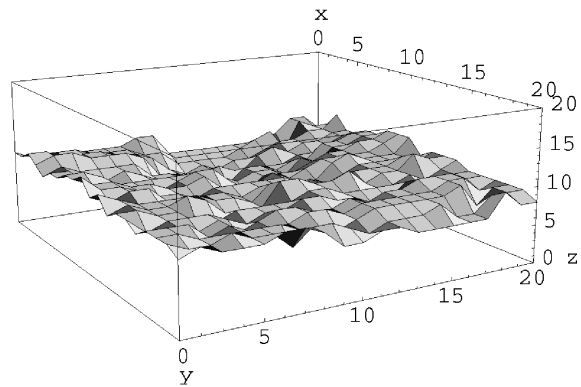}
  \caption{One of the two interfaces of the slab state on a cubic lattice of
    linear dimension $20$.  Coordinates are in units of lattice spacing.}
\label{slab}
\end{figure}

The main result of the above analysis is that the average consensus time is
much larger than the dependence of $N^{2/d}$ that would arise if domain
coarsening were entirely governed by diffusive dynamics.  By observing the
evolution of many realizations of the system, it is clear that the asymptotic
tail of the consensus time distribution arises from situations where the $+$
and $-$ spins organize into spatially coherent and long-lived states that
consist of relatively flat stripes in two dimensions (Fig.~\ref{pics}), slabs
in three dimensions (Fig.~\ref{slab}), and analogously (we believe) in higher
dimensions.  The existence of these states is one of the most surprising
feature of the MR model.  In spite of the isotropy of the MR interaction, the
long-lived transient states arise and spontaneously break this symmetry.
Once the system reaches such a state, further evolution proceeds extremely
slowly, as we shall discuss below.

To develop intuition for these coherent states, we show in Fig.~\ref{pics} a
set of snapshots of a $50\times 50$ system that happens to evolve to a
stripe.  After a few time steps, the lattice-scale granularity of the random
initial state has disappeared due to the effective surface tension in the
majority rule dynamics.  After this early-time transient, the subsequent
evolution qualitatively resembles the coarsening of a spin system with
non-conserved order-parameter kinetics.  However, domains tend to develop a
stringy morphology, a feature that promotes the formation of stripes that
span the system.  For the realization shown, a clearly resolved stripe
emerges by 100 time steps, while ultimate consensus is achieved when 1850
time steps have elapsed; notice that a time of 1850 steps is relatively early
in the asymptotic tail of $P_N(t)$ in Fig.~\ref{PDF}.

In spite of the anomalous long-time kinetics of the MR model, the early-time
coarsening is diffusive in nature.  To determine the growth of the typical
domain length scale at early times, we studied the time evolution of the
two-spin correlation function.  We took this correlation function at
different times and found the length rescaling that gave the best data
collapse.  We thus found that the appropriate rescaling the correlation
function is by a length scale that is proportional to $t^{1/2}$.  We
therefore conclude that the early-time coarsening in the MR model is
characterized by a length scale that grows as $t^{1/2}$.

A phenomenon analogous to stripe formation occurs in three dimensions, where
long-lived states arise that consist of two relatively flat slabs of
oppositely-oriented spins (Fig.~\ref{slab}).  For the example shown from a
$20^3$ lattice, a slab state forms around 150 time steps, while final
consensus does not occur until 3800 time steps have elapsed.

\begin{figure}[ht] 
 \vspace*{0.cm}
 \includegraphics*[width=0.425\textwidth]{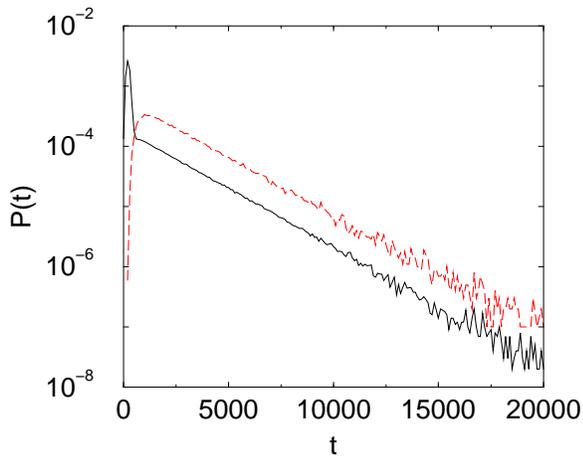}
 \caption{Consensus time distribution for a $50^2$ square lattice for a
   two-stripe (dashed) and random initial condition (solid).  All data are
   based on $10^5$ realizations.  Time is in units of Monte Carlo steps per
   spin.}
\label{stripe+random}
\end{figure}

To verify that stripe states actually govern the asymptotic tail of $P_N(t)$
two dimensions, we also study the evolution of a synthetic system with an
ordered initial state that consists of two straight stripes, with half the
spins $+$ and half the spins $-$.  The long-time tail of the consensus time
distribution for this special initial condition follows a single
exponentially decaying function, as shown in Fig.~\ref{stripe+random}.  Also
shown in this figure is the corresponding distribution for a system of the
same size with a random zero-magnetization initial condition.  The
coincidence of the slopes in the tails of these two distributions shows that
stripe states control the long-time evolution of random zero-magnetization
initial condition systems.

Because of the crucial role that spatially coherent states play in the MR
model, we also study the probability $S_N(m)$ that a randomly-prepared
$N$-spin system in $d$ dimensions with initial magnetization $m$ evolves to
such a state.  We use two independent methods to measure $S_N(m)$.  One is
based on simply counting the fraction of realizations whose consensus time
lies within the asymptotic tail of the consensus time distribution.  For
example, for the data from the $50\times 50$ system in Fig.~\ref{PDF}, the
tail region corresponds to a consensus time $t$ greater than 600.  Thus for
this system size all realizations with $t>600$ are counted as reaching a
coherent state.

Alternatively, we investigate correlation functions that are engineered to
detect stripe states.  For $d=2$, we consider the following correlation
functions for two spins that are located a distance $L/2$ apart:
\begin{eqnarray*}
C_x(t)&\equiv&\frac{1}{2} \langle s(x,y,t)s(x\pm L/2,y,t)\rangle, \\
C_y(t)&\equiv&\frac{1}{2} \langle s(x,y,t)s(x,y\pm L/2,t)\rangle.
\end{eqnarray*}
For both a random state and for consensus, these correlation functions equal
zero.  Conversely, for an ordered two-stripe state with stripes of width
$L/2$ parallel to the $x$-axis, $C_x=+1$ and $C_y=-1$, and vice versa for
stripes parallel to the $y$-axis.  We therefore posit that a stripe state
arises if the correlation function in the direction(s) parallel to the stripe
is greater than a threshold value, while the correlation function
perpendicular to the stripe is less than the negative of this threshold
value.  We arbitrarily choose the threshold to equal 0.5, but our results for
large $N$ depend only weakly ($S_N(m=0)$ varies by $\leq10\%$) on the
threshold value when it is in the range 0.3 -- 0.7.  The results given below
are based on the threshold set to 0.5.

\begin{figure}[ht] 
 \vspace*{0.cm}
 \includegraphics*[width=0.425\textwidth]{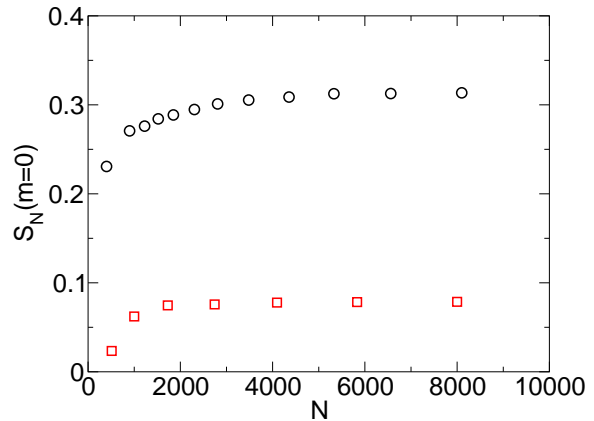}
 \caption{Probability to reach a stripe state versus number of spins $N$ for
   $d=2$ ($\circ$) and $d=3$ ($\Box$) using the threshold value 0.5 (see
   text).  Data are based on $10^5$ realizations.}
\label{psvsn}
\end{figure}

\begin{figure}[ht] 
 \vspace*{0.cm}
 \includegraphics*[width=0.425\textwidth]{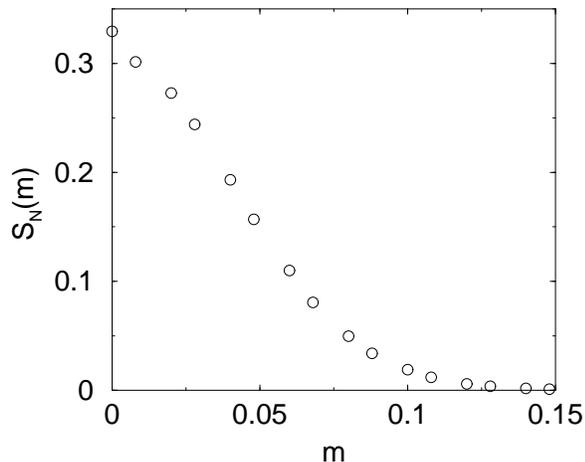}
 \caption{Probability to reach a stripe state versus initial magnetization
   $m$ for a $50^2$ system.  Data are based on $10^5$ realizations.}
\label{2d_psvsx}
\end{figure}

We find that the stripe/slab probability $S_N(m=0)$ grows quickly for small
$N$ and then saturates to a non-zero value that is close to 0.33 in $d=2$ and
0.08 for $d=3$ (Fig.~\ref{psvsn}).  The stripe probability in two dimensions
is very close to that found previously in the zero-temperature evolution of
the Ising model with Glauber kinetics \cite{SKR}.  Note also that as the
initial magnetization $m$ is moved away zero, $S_N(m)$ quickly decays to zero
(Fig.~\ref{2d_psvsx}).  This simply reflects the fact that if one phase is
initially below the percolation threshold, there is a very small possibility
for minority phase droplets to merge and form a stripe that spans the system.

We can qualitatively understand the dimension dependence of the probability
to reach a stripe state by the following rough argument (see also
Ref.~\cite{SKR}).  For simplicity, we first discuss the case of two
dimensions with the random zero-magnetization initial condition.  Consider a
large system of linear dimension $L$ and cut it into four equal subsquares of
linear size $L/2$.  The final state in each of these subsquares is reached
more quickly than that of the entire system.  We now make the plausible
assumption, based on observations of many realizations of the system, that
each subsquare independently reaches consensus.  Then out of the $2^4$
possible configurations of these subsquares, only the following arrangements
\begin{eqnarray*}
\begin{array}{llll}
~~++ &  ~~-- & ~~+- & ~~-+ \\
~~-- &  ~~++ & ~~+- & ~~-+,
\end{array}
\end{eqnarray*}
where the $+$ and $-$ symbols refer to the final state of each subsquare,
correspond to a stripe state of the $L\times L$ system.  This argument then
suggests that $S_N(m=0)=4/2^4=1/4$.

This coarse-graining argument straightforwardly generalizes to higher
spatial dimensions.  On the cubic lattice, we divide an $L\times L\times L$
cube into eight subcubes of linear dimension $L/2$.  If these subcubes each
independently reach consensus, then a slab state on the original cube
(consisting of two slabs of oppositely oriented spins, each of size $L\times
L\times L/2$) can be achieved in six possible ways.  The probability of
reaching a slab state is therefore $6/2^8\approx 0.047$.  In $d$ dimensions,
this same line of reasoning gives $S_N(m=0)= 2d/2^{2^d}$.  While our argument
is crude, the resulting numerical values for $S_N$ qualitatively mirror the
corresponding estimates from simulations.

Our approach also helps explain why stripe states quickly disappear when the
initial magnetization is non-zero.  As an example, for initial magnetization
0.08, we find by numerical simulations that the probability that a $25\times
25$ system eventually ends with all spins $+$ is 0.88.  Now employing the
above coarse-graining argument for a $50\times 50$ system, the four $25\times
25$ subsquares will each reach $+$ consensus with probability 0.88 and $-$
consensus with probability 0.12.  Then the probability for the $50\times 50$
system to reach a stripe state is $S_N\approx 4 (0.88)^2\, (0.12)^2 \approx
0.0446$.  This is very close (probably fortuitously) with our numerical
result of $S_N(m=0.08)\approx 0.0498$.

\begin{figure}[ht] 
 \vspace*{0.cm}
 \includegraphics*[width=0.45\textwidth]{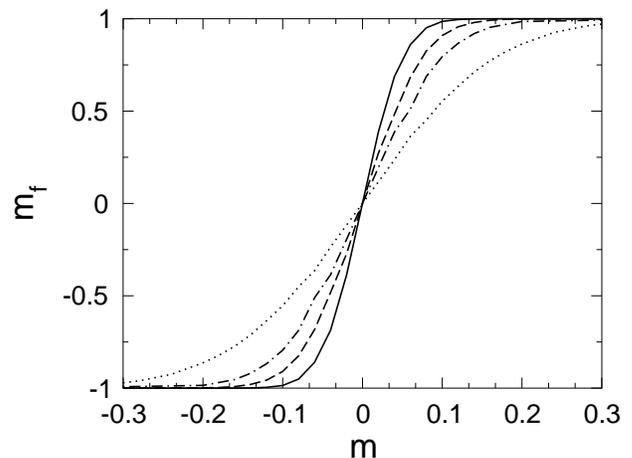}
 \caption{Final magnetization $m_f$ as a function of initial magnetization
   $m$, on the square lattice for linear dimension $L=10$ (dotted curve), 20
   (dashed), 30 (dot-dash), and 50 (solid).}
\label{2d_mfvsmi}
\end{figure}

Another important aspect of the evolution to the final state is the
dependence of the final magnetization on the initial magnetization.  Since
the system always reaches consensus, the final magnetization $m_f$ is simply
the difference in the probabilities that the systems ends with all spins $+$
and all spins $-$.  On the square lattice, we find that the curve of the
final magnetization versus the initial magnetization approaches a step
function as $N\to\infty$ (Fig.~\ref{2d_mfvsmi}).  Thus any initial bias
pre-determines the final state of the system in the thermodynamic limit.
This step-function behavior is in contrast to the behavior in one dimension,
where the final magnetization curve remains non-singular as $N \rightarrow
\infty$ \cite{MM}.  Finally, it is worth noting that $m_f$ equals the initial
magnetization $m$ for the voter model in all spatial dimensions \cite{voter};
there is no tyranny of the majority in the voter model.

\section{Lifetime of Stripe States}

Once a stripe state is formed, the evolution to ultimate consensus is
controlled by the time required for the two interfaces that define the stripe
to meet and annihilate.  In one dimension, it is easy to see that each
isolated interface between $+$ and $-$ spins moves by free diffusion.  When
two interfaces approach to nearest-neighbor separation they necessarily
annihilate.  (Note that in the kinetic Ising model, two nearest-neighbor
interfaces can annihilate, with probability 1/2, or recede by one lattice
spacing, also with probability 1/2, in a single update step.)~ Therefore the
time for the last two domain walls to annihilate is proportional to $L^2$,
where $L$ is the linear dimension of the system.  Further, because the system
is controlled by the meeting of two random walks on a finite ring, the
consensus time distribution has an exponential decay of the form $e^{-t/N^2}$
\cite{fpp}.

In two and three dimensions, the interfaces between stripes are quite smooth
(Figs.~\ref{pics} \& \ref{slab}) and the scaling of the interface width on
the transverse dimension of the system appears to be in the Edwards-Wilkinson
universality class \cite{BS}.  We verified the smoothness of the interface by
preparing a system of linear size ${\cal L}\times L^{(d-1)}$ in $d$
dimensions, with ${\cal L}\gg L$, in which all spins in the region $[0,{\cal
  L}/2]$ are initially in the $-$ state, and all spins in the region $[{\cal
  L}/2,{\cal L}]$ are initially in the $+$ state.  In two dimensions, the
width $w$ of the interface initially grows slowly in time and eventually
saturates to a value that approximately scales as $w\sim L^{1/2}$.  In three
dimensions, the growth of the width is even slower and the saturation value
of the width is consistent with a logarithmic dependence on $L$.

Thus it is the diffusion of the interface as a whole rather than fluctuations
in the interface shape that determines the lifetime of the stripe state.
Since the interfaces are typically separated by a distance of order $L$ when
they are first formed, the lifetime $T$ of the stripe state should therefore
given by
\begin{equation}
\label{ew_2d}
T \sim \frac{L^2}{D(L)},
\end{equation} 
where $D(L)$ is the diffusion coefficient of a single interface with
transverse dimension $L$.

We may obtain a simple albeit rough estimate for this diffusion coefficient
by treating each site on the interface as an independent random walk
\cite{SKR,PRL}.  For a $d$-dimensional system, a smooth interface contains of
the order of $L^{d-1}$ sites.  In a single time step, each interface site
will randomly move by $\pm 1$ perpendicular to the interface.  Hence if each
site is independent, the center of mass of the interface will move by a
distance $\sqrt{L^{d-1}}/L^{d-1}\sim L^{-(d-1)/2}$ in one time step.  As a
result, the diffusion coefficient of the interface $D(L)$ scales as
$L^{-(d-1)}$.

We tested this prediction by simulation by following the evolution of a
single interface in a long strip (or slab) geometry with transverse dimension
$L$ in which all the spins on the right half are set to $+1$ and all the
spins on the left half are set to $-1$.  We then let the spins evolve by
majority rule dynamics.  After a short transient that lasts of the order of
one time step, we observe that the interface moves diffusively, with a
diffusion coefficient that scales approximately as $L^{-1}$ in two dimensions
and as $L^{-2}$ in three dimensions.  Given the crudeness of the above random
walk argument, it is surprising that the simulation results agree quite well
with the prediction $D(L)\sim L^{-(d-1)}$.

From this scaling of the diffusion coefficient on the transverse linear
dimension $L$, Eq.~(\ref{ew_2d}) then gives a consensus time $T$ that scales
as $T\sim L^{d+1}$.  Equivalently, in terms of the total number of spins
$N=L^d$, the dependence is $T\sim N^{(d+1)/d}$.  However, this prediction is
only qualitatively consistent with the exponent values of $1.7$ for $d=2$ and
$1.5$ for $d=3$ that were obtained from direct numerical simulations of the
consensus time distribution.  We do not have an explanation for this
discrepancy.

\section{Summary and Discussion}

We studied the time evolution of the majority rule (MR) model for
finite-dimensional systems.  One of our main results is that the approach to
consensus in an initially unbiased system is surprisingly complex.  Before
ultimate consensus is reached, a non-trivial fraction of all realizations
falls into coherent metastable states that consist of stripes in two
dimensions and slabs in three dimensions.  We anticipate that analogous
coherent states arise in higher spatial dimensions.  The interfaces between
domains in these coherent states are quite smooth and reflect the strong
surface tension in the majority rule dynamics.

\begin{figure}[ht] 
 \vspace*{0.cm}
 \includegraphics*[width=0.425\textwidth]{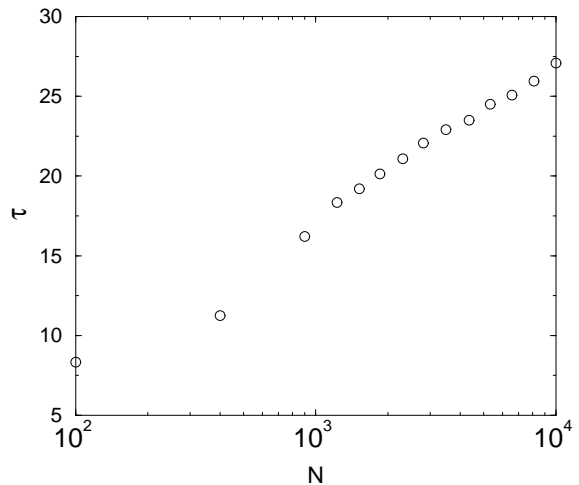}
 \caption{Consensus time (in units of Monte Carlo steps per spin) versus $N$
   in $d=2$ for initial magnetization $m=0.2$.}
\label{t-asym}
\end{figure}

Due to these coherent states, the time to reach consensus is anomalously long
and is controlled by a diffusion process that brings two interfaces close
enough that they can annihilate.  The characteristic time scale for this
annihilation is much longer than the most probable time to reach consensus.
The fraction of realizations that reach these long-lived states decreases as
$d$ increases, but their role appears to be dominant in the asymptotic
kinetics.  When the initial magnetization is non-zero, however, these
long-lived states quickly disappear.  As a result, the consensus time
distribution has a single peak and there is no long-time tail.  Furthermore,
the time until consensus grows only logarithmically in the system size
(Fig.~\ref{t-asym}).  Thus an initial bias in the density of spins is a
decisive influence in the long-time behavior of the system.

We gave a crude coarse-graining argument to estimate the probability to reach
a coherent state as a function of the spatial dimension.  This approach
qualitatively explained the behavior of the probability to reach such a state
as a function of the spatial dimension $d$ and the initial magnetization.
 
Finally, we suggest several directions for further study.  First, it would be
worthwhile to determine the value of the upper critical dimension of the MR
model.  An exact analysis of this model on the complete graph, where all
spins are nearest neighbors of each other, showed that the mean consensus
time grows as $\ln N$ \cite{MM}.  On the other hand, the simulations
presented here and in \cite{MM} suggested that the mean consensus time grows
as a power law in $N$ for spatial dimensions 1, 2, 3, and 4.  These two facts
suggest that the upper critical dimension of the MR model is greater than 4.
It would be worthwhile to have a theoretical understanding for the apparently
large value of the upper critical dimension.

We also believe it will be fruitful to study simple extensions of the MR
model with more stringent conditions for achieving consensus.  One example is
to have a higher threshold than simple majority before the opinion of a group
is swayed.  While a higher threshold will obviously slow the dynamics, it
should be interesting to investigate whether this modification leads to
different scaling properties for the mean consensus time and the distribution
of consensus times.

A more intriguing generalization arises when each spin has more than two
opinions, where we anticipate new types of dynamical behavior.  With more
than two states, the possibility of a dynamically-stable steady state that
consists of coalescing and coexisting multiple opinion groups was discussed
in the framework of the ``stochastic seceder'' model \cite{SH03}.  In the
context of our majority rule model, there obviously will be slower dynamics
because it may be possible to have a group with no local majority, but only a
local plurality.  Such a group would not evolve according to the majority
rule dynamics.  Thus configurations in which there is no majority in each
group represent another absorbing state for the dynamics.  In the mean-field
limit, we find that such a system never reaches this frustrated state, as the
corresponding fixed point of equal concentrations of all species is unstable
\cite{CR}.  Instead, for a system with more than two opinion states, the time
to reach ultimate consensus is merely increased by a multiplicative factor
compared to the two-opinion MR model.  However, for finite spatial
dimensions, the existence of more than two opinions appears to have a more
significant effect on the long-time behavior that depends fundamentally on
the interplay between the group size and the number of states.  When there
are many distinct local majorities in the initial state the group dynamics
has a primarily diffusive character.  However, when there is of the order of
one local majority, the opinion of this group quickly overtakes the entire
system.

\section{Acknowledgments}

We thank Pablo Hurtado, Paul Krapivsky, Mauro Mobilia, and Federico Vazquez
for helpful discussions and advice.  We also thank NSF grant DMR0227670 for
financial support of this research.

\end{document}